\documentclass{article}

\usepackage{cite}
\usepackage{graphicx}
\usepackage{dcolumn}

\begin{document}

\date{}
\title{Accurate eigenvalues of the Schr\"{o}dinger equation with the potential
$V(r)=V_{0}r^{\alpha }$}
\author{Francisco M. Fern\'{a}ndez \thanks{%
E-mail: fernande@quimica.unlp.edu.ar} \\
%EndAName
INIFTA (CONICET, UNLP), Divisi\'on Qu\'imica Te\'orica\\
Blvd. 113 S/N, Sucursal 4, Casilla de Correo 16, 1900 La Plata, Argentina}
\maketitle

\begin{abstract}
We calculate accurate eigenvalues of the Schr\"{o}dinger equation with the
potential $V(r)=V_{0}r^{\alpha }$, $\alpha \geq -1$, $V_{0}\alpha >0$. We
resort to the Riccati-Pad\'{e} method that is based on a rational
approximation to the logarithmic derivative of the wavefunction. This
approach applies when $\alpha $ is a rational number.
\end{abstract}

\section{Introduction}

\label{sec:intro}

Some time ago Fern\'{a}ndez et al\cite{FGZ93} proposed a
modification of the Riccati-Pad\'{e} method
(RPM)\cite{FMT89a,FMT89b} for the calculation of the eigenvalues
of the Schr\"{o}dinger equation with the potential
$V(r)=gr^{\alpha }$, where $\alpha >-2$ is a rational exponent and
$g\alpha
>0 $. As an illustrative example they applied the standard RPM to
the case $\alpha =-1/2$ and obtained the ground-state energy quite
accurately\cite {FGZ93}.

In a recent paper Li and Dai\cite{LD16} showed that the Schr\"{o}dinger
equation with the potential $V(r)=-\alpha r^{-1/2}$, $\alpha >0$, can be
solved exactly in terms of Heun biconfluent functions. They obtained the
eigenvalues $E_{\nu ,l}$ for $\nu ,l=0,1,\ldots ,5$, where $\nu $ and $l$
are the well known radial and angular-momentum quantum numbers,
respectively. In particular, their estimate $E_{00}=-0.4380$ agrees with the
more accurate RPM result $E_{00}=-0.438041241942506$\cite{FGZ93}.

The purpose of this paper is to show that the standard RPM\cite
{FMT89a,FMT89b} (see also \cite{F08} for more details and references) is a
suitable tool for the accurate calculation of the eigenvalues of the
potential $V(r)=V_{0}r^{\alpha }$, where $\alpha \geq -1$ is a rational
number and $V_{0}\alpha >0$. In section~\ref{sec:RPM} we outline the
approach, in section~\ref{sec:examples} we apply it to selected examples and
discuss the accuracy of the results and in section~\ref{sec:conclusions} we
draw conclusions.

\section{The Riccati-Pad\'{e} method}

\label{sec:RPM}

The Schr\"{o}diger equation for the present central-field model is
\begin{eqnarray}
H\psi &=&E\psi ,  \nonumber \\
H &=&-\frac{\hbar ^{2}}{2m}\nabla ^{2}+V_{0}r^{\alpha },\;\alpha V_{0}>0.
\end{eqnarray}
On choosing the units of length $r_{0}=\left[ \hbar ^{2}/\left( 2m\left|
V_{0}\right| \right) \right] ^{1/(\alpha +2)}$ and energy $e_{0}=\left[
\hbar ^{2}\left| V_{0}\right| ^{1/(\alpha +1)}/(2m)\right] ^{(\alpha
+1)/(\alpha +2)}$ the resulting dimensionless Hamiltonian becomes
\begin{equation}
H=-\nabla ^{2}+\sigma r^{\alpha },\;\sigma =\frac{\alpha }{\left| \alpha
\right| }=\frac{V_{0}}{\left| V_{0}\right| }.
\end{equation}

The solutions are of the form $\psi (r,\theta ,\phi )=R(r)Y_{l}^{m}(\theta
,\phi )$, where $Y_{l}^{m}(\theta ,\phi )$ are the spherical harmonics with
angular-momentum quantum numbers $l=0,1,\ldots $ and $m=0,\pm 1,\ldots ,\pm
l $. The radial part of the solution satisfies the eigenvalue equation
\begin{equation}
-u^{\prime \prime }(r)+\left[ \frac{l(l+1)}{r^{2}}+\sigma r^{\alpha }\right]
u(r)=\epsilon u(r),
\end{equation}
where $u(r)=rR(r)$ and $\epsilon =E/e_{0}$ is the dimensionless energy.

The modified logarithmic derivative
\begin{equation}
f(r)=\frac{l+1}{r}-\frac{u^{\prime }(r)}{u(r)},
\end{equation}
satisfies the Riccati equation
\begin{equation}
f^{\prime }(r)=f(r)^{2}-\frac{2(l+1)}{r}f(r)+\epsilon -\sigma r^{\alpha }.
\end{equation}

In order to apply the RPM we define the new independent and dependent
variables $z=r^{\beta }$ and $g(z)=f(r(z))$, respectively; the latter
satisfies the Riccati equation
\begin{equation}
\beta zg^{\prime }(z)+2(l+1)g(z)=z^{1/\beta }g(z)^{2}+\epsilon z^{1/\beta
}-\sigma z^{(\alpha +1)/\beta }.  \label{eq:Riccati_g_a}
\end{equation}
If $\alpha =p/q$, where $p$ and $q$ are integers, we choose $\beta =1/q$ so
that the Riccati equation (\ref{eq:Riccati_g_a}) becomes
\begin{equation}
\frac{1}{q}zg^{\prime }(z)+2(l+1)g(z)=z^{q}g(z)^{2}+\epsilon z^{q}-\sigma
z^{p+q}.  \label{eq:Riccati_g_b}
\end{equation}
If $q>0$ and $p+q\geq 0$ (which lead to $\alpha \geq -1$) then we can expand
the solution of (\ref{eq:Riccati_g_b}) in a Taylor series about $z=0$
\begin{equation}
g(z)=\sum_{j=0}^{\infty }g_{j}z^{j},
\end{equation}
where the coefficients $g_{j}$ are polynomial functions of $\epsilon $. They
can be obtained from the recurrence relation
\begin{equation}
g_{n}=\frac{1}{2l+\frac{n}{q}+2}\left[
w(n-q)\sum_{j=0}^{n-q}g_{j}g_{n-q-j}+\epsilon \delta _{nq}-\sigma \delta
_{n\,p+q}\right] ,\;n=0,1,\ldots ,  \label{eq:g_n_rec_rel}
\end{equation}
where $w(x)$ is the Heaviside function ($w(x)=0$ if $x<0$ and $w(x)=1$
otherwise).

If we look for a rational approximation to the solution of (\ref
{eq:Riccati_g_b}) of the form
\begin{equation}
\left[ M,N\right] (z)=\frac{\sum_{i=0}^{M}a_{i}z^{i}}{1+%
\sum_{j=1}^{N}b_{j}z^{j}}=\sum_{k=0}^{M+N+1}g_{k}z^{k}+\mathcal{O}\left(
z^{M+N+2}\right) ,  \label{eq:Padé}
\end{equation}
then the approximate eigenvalue $\epsilon $ should be a root of the Hankel
determinant $H_{D}^{d}(\epsilon )=0$, where $D=N+1$ and $d=M-N$. The matrix
elements of this determinant are $g_{i+j+d-1}$, $i,j=1,2,\ldots ,D$. Earlier
applications of the RPM showed that there are sequences of roots $\epsilon
^{[D,d]}$, $D=2,3,\ldots $ that converge towards the actual eigenvalues of
the problem\cite{FMT89a,FMT89b} (in particular, see \cite{F08} and
references therein). Typically, the rate of convergence exhibits exponential
behaviour $\left| \epsilon ^{[D+1,d]}-\epsilon ^{[D,d]}\right| =Ae^{-BD}$
for sufficiently large $D$.

When both $q$ and $p+q$ are odd, then $g(z)$ is odd and $g_{2k}=0$, $%
k=0,1,\ldots $. Although equation (\ref{eq:g_n_rec_rel}) and the
prodecure just outlined are still valid under these conditions it
only makes sense to construct the sequences of roots with either
$D$ even or $D$ odd or, which is more convenient from a practical
point of view, to construct the Hankel determinants directly from
the nonzero expansion coefficients $g_{2j+1}$
instead of $g_{j}$. Note that the actual expansion variable in this case is $%
z^{2}$ instead of $z$.

\section{Examples}

\label{sec:examples}

In this section we apply the RPM to several examples. We calculate the
expansion coefficients $g_{j}$ and the Hankel determinants $H_{D}^{d}$
analytically as polynomial functions of $\epsilon $ and then obtain the
roots of $H_{D}^{d}(\epsilon )=0$ numerically. Although the rate of
convergence may slightly vary with the chosen value of $d$ we restrict
ourselves to the case $d=2$ for concreteness. It is well known that the RPM
yields the actual eigenvalues disregarding the asymptotic behaviour of the
selected rational function $[M,N](z)$\cite{FMT89a,FMT89b,F08}. The
calculation of the eigenvalues by means of the RPM is quite straightforward
as it reduces to finding convergent sequences of roots $\epsilon ^{[D,d]}$, $%
D=2,3,\ldots $, of the polynomial equation  $H_{D}^{d}(\epsilon )=0$. In
what follows we label the estimated dimensionless energies as $\epsilon
_{l\,\nu }$, where $l,\nu =0,1,\ldots $ are the angular momentum and radial
quantum numbers, respectively.

The first example is given by the exponent $\alpha =-1/2$ that was treated
earlier by means of the RPM\cite{FGZ93} and has recently been proved to lead
to a Schr\"{o}dinger equation that is exactly solvable in terms of Heun
biconfluent functions\cite{LD16}. Here we apply the standard RPM outlined in
section~\ref{sec:RPM} with $p=-1$ and $q=2$.

\begin{figure}[tbp]
\caption{Logarithmic error $L_D$ for the eigenvalues of
$V(r)=-r^{-1/2}$ with $l=0$ and $n=0,1,2,3,4,5$ (from left to
right)} \label{fig:rm12l0}
\begin{center}
\includegraphics[width=9cm]{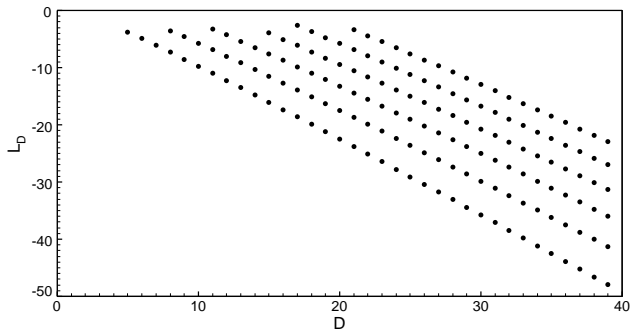}
\end{center}
\end{figure}

Figure~\ref{fig:rm12l0} shows the exponential rate of convergence in terms
of the logarithmic error $L_{D}=\log \left| \epsilon _{0\,\nu
}^{[D+1,2]}-\epsilon _{0\,\nu }^{[D,2]}\right| $ for $D_{\nu }\leq D\leq 40$%
. We appreciate that the slope of $L_{D}$ vs $D$ is almost independent of $%
\nu $ but the starting point $D_{\nu }$ of each sequence increases with $\nu
$. It means that a given accuracy is obtained with increasingly greater
determinant dimension as $\nu $ increases. The reason is that the number of
nodes of $u(r)$ increases with $\nu $ and the degree $N$ of the polynomial
in the denominator of the Pad\'{e} approximant (\ref{eq:Padé}) should
increase accordingly. Exactly the same situation takes place for all $l>0$.

Table~\ref{tab:rm12} shows some eigenvalues estimated from the sequences of
roots of the Hankel determinants. The error is supposed to be in the last
digit. The first four digits of present RPM eigenvalues agree with those
obtained from the analytical expressions for the bound-state eigenfunctions%
\cite{LD16}.

Table~\ref{tab:rm13} shows results for $\alpha =-1/3$ obtained from Hankel
determinants of dimension $D\leq 45$. We appreciate that the rate of
convergence is slightly smaller than for the preceding example.

Table~\ref{tab:rm23} shows results for $\alpha =-2/3$ where the Hankel
determinants of dimension $D\leq 40$ were constructed from the coefficients $%
g_{2j+1}$ as indicated in section~\ref{sec:RPM}. At present we do not know
the reason for the remarkable rate of convergence clearly observed in this
particular case.

Tables \ref{tab:r12}, \ref{tab:r13}, \ref{tab:r23} and \ref{tab:r32} show
results for $\alpha =1/2$ ($D\leq 45$), $\alpha =1/3$ ($D\leq 45$), $\alpha
=2/3$ ($D\leq 40$) and $\alpha =3/2$ ($D\leq 45$), respectively. In the case
$\alpha =2/3$ the Hankel determinants were constructed from the coefficients
$g_{2j+1}$. In all these cases the results are also quite accurate due to
the exponential rate of convergence.

\section{Conclusions}

\label{sec:conclusions}

The Schr\"{o}dinger equation with the potential $V(r)=-r^{-1/2}$ has been
shown to be solvable in terms of known functions\cite{LD16}. However, the
resulting quantization condition does not seem quite amenable for the
accurate calculation of the eigenvalues. This fact motivated us to apply the
RPM to this problem as well to other ones with potential-energy functions of
the same general form. In this way we considerably extended the calculations
carried out in a previous application of the RPM to this kind of
quantum-mechanical models\cite{FGZ93}. Present accurate results may be a
useful benchmark for testing other numerical approaches.

\begin{table}[tbp]
\caption{Some eigenvalues for the potential $V(r)=-r^{-1/2}$}
\label{tab:rm12}
\begin{center}
\par
\begin{tabular}{cD{.}{.}{48}}
\hline \multicolumn{1}{c}{$(l,\nu)$}&
\multicolumn{1}{c}{$\epsilon_{l\,\nu}$}  \\
\hline
$(0,0)$ & -0.438041241942505887099625301211018073928496394488 \\
$(0,1)$ & -0.263203069697058806981126460451511573625086 \\
$(0,2)$ & -0.197558399925620717779064487394853997 \\
$(0,3)$ & -0.16170496623669019780841784885683  \\
$(0,4)$ & -0.138637391239231884846885804       \\
$(0,5)$ & -0.12234576385330475915767             \\
$(1,0)$ & -0.2866109687201626908136243251996742333626644870  \\
$(1,1)$ & -0.2098001468533432750929194383080363872292       \\
$(1,2)$ & -0.16941598774824196934573345654571593        \\
$(1,3)$ & -0.144018998103941636732449264637            \\
$(1,4)$ & -0.12635429124801362717115287            \\
$(1,5)$ & -0.1132459841978892683222             \\
$(2,0)$ & -0.2215058763914675647366137657430051657462702    \\
$(2,1)$ & -0.176817135425491179207487464178593305496   \\
$(2,2)$ & -0.1491980852730719082193440172127977     \\
$(2,3)$ & -0.130219955532154314232404909998           \\
$(2,4)$ & -0.11626314686784459628268595              \\
$(2,5)$ & -0.1055030282421431483474                 \\
$(3,0)$ & -0.1840051303223707035150871277288602361291698   \\
$(3,1)$ & -0.15423877223980968359730523226455785475       \\
$(3,2)$ & -0.1339881376313473234199035428225626       \\
$(3,3)$ & -0.11920777050369868492801424398           \\
$(3,4)$ & -0.1078802665413018675350206          \\
$(3,5)$ & -0.988821792686294467534              \\

\end{tabular}
\par
\end{center}
\end{table}

\begin{table}[tbp]
\caption{Some eigenvalues for the potential $V(r)=-r^{-1/3}$}
\label{tab:rm13}
\begin{center}
\par
\begin{tabular}{cD{.}{.}{30}}
\hline \multicolumn{1}{c}{$(l,\nu)$}&
\multicolumn{1}{c}{$\epsilon_{l\,\nu}$}  \\
\hline
$(0,0)$ & -0.5497449679398553680821393889 \\
$(0,1)$ & -0.4024649710757823079445 \\
$(0,2)$ & -0.3381558345859440423 \\
$(1,0)$ & -0.42871166425372295840404887  \\
$(1,1)$ & -0.3533007381052660764965       \\
$(1,2)$ & -0.30983106016546652        \\
$(2,0)$ & -0.36806218373376176661841329    \\
$(2,1)$ & -0.28793956847797973   \\
$(3,0)$ & -0.329645913475774067101662   \\
$(3,1)$ & -0.2704993104962070       \\

\end{tabular}
\par
\end{center}
\end{table}

\begin{table}[tbp]
\caption{Some eigenvalues for the potential $V(r)=-r^{-2/3}$}
\label{tab:rm23}
\begin{center}
\par
\begin{tabular}{cD{.}{.}{70}}
\hline \multicolumn{1}{c}{$(l,\nu)$}&
\multicolumn{1}{c}{$\epsilon_{l\,\nu}$}  \\
\hline
$(0,0)$ & -0.3553826960845427527963094045643964777566071121201478986602873029449984 \\
$(0,1)$ & -0.16868325808153151960154553282980972498667144444080356922104021 \\
$(0,2)$ & -0.11038365898943484824621298400145968475548008298916638897   \\
$(0,3)$ & -0.081999526908911404704158808703005121437968376047937  \\
$(0,4)$ & -0.06521809931436284998665444067267017535170570716       \\
$(0,5)$ & -0.0541352674541698627240411071289680985899564        \\
$(0,6)$ & -0.04627062511812482630789151371897668129    \\
$(0,7)$ & -0.404005371138502882697365370913864   \\
$(0,8)$ & -0.0358518031538109882986676433247   \\
$(0,9)$ & -0.032223489306250183645301254       \\
$(1,0)$ & -0.1850179056602088890332800229212408125011697981207067554954681974224         \\
$(1,1)$ & -0.117968431924077608469524665145254831699685327441035531300195  \\
$(1,2)$ & -0.0863759461243326673189313655538476113508242491487887433   \\
$(1,3)$ & -0.068067125029383214620224044370102299284619470137459 \\
$(1,4)$ & -0.0561383821480827319457487775745597084348703163    \\
$(1,5)$ & -0.04775631353700811583226913054057711724803  \\
$(1,6)$ & -0.036762960762327517031166177241246  \\
$(1,7)$ & -0.02987774967642271810296609  \\

\end{tabular}
\par
\end{center}
\end{table}

\begin{table}[tbp]
\caption{Some eigenvalues for the potential $V(r)=r^{1/2}$}
\label{tab:r12}
\begin{center}
\par
\begin{tabular}{cD{.}{.}{30}}
\hline \multicolumn{1}{c}{$(l,\nu)$}&
\multicolumn{1}{c}{$\epsilon_{l\,\nu}$}  \\
\hline
$(0,0)$ & 1.833393609778132819989706566163148984 \\
$(0,1)$ & 2.5506474914147899630015973569   \\
$(0,2)$ & 3.051181948950148127778090       \\
$(0,3)$ & 3.4521319438575226071            \\
$(0,4)$ & 3.7933604446476296              \\
$(1,0)$ & 2.300496239515583918636898196682148  \\
$(1,1)$ & 2.854335925747438604342984490    \\
$(1,2)$ & 3.285833295818405737010872         \\
$(1,3)$ & 3.64738542145476544982        \\
$(1,4)$ & 3.9626765006936562         \\
$(2,0)$ & 2.6575633682836919447122189696955    \\
$(2,1)$ & 3.12032849206100807933805775         \\
$(2,2)$ & 3.50245154742884623274796        \\
$(2,3)$ & 3.8325439158453068899         \\
$(2,4)$ & 4.125809074413673        \\
$(3,0)$ & 2.9544509310223920360300214507054  \\
$(3,1)$ & 3.70270499761415995593668     \\
$(3,2)$ & 4.0073673396275475504      \\
$(3,3)$ & 4.2819594417211         \\

\end{tabular}
\par
\end{center}
\end{table}

\begin{table}[tbp]
\caption{Some eigenvalues for the potential $V(r)=r^{1/3}$}
\label{tab:r13}
\begin{center}
\par
\begin{tabular}{cD{.}{.}{30}}
\hline \multicolumn{1}{c}{$(l,\nu)$}&
\multicolumn{1}{c}{$\epsilon_{l\,\nu}$}  \\
\hline
$(0,0)$ & 1.61567508878829338502205   \\
$(0,1)$ & 2.04183293233151969   \\
$(0,2)$ & 2.319639064734      \\
$(1,0)$ & 1.90488667402162636938799 \\
$(1,1)$ & 2.21665869399853924      \\
$(1,2)$ & 2.448777374655        \\

\end{tabular}
\par
\end{center}
\end{table}

\begin{table}[tbp]
\caption{Some eigenvalues for the potential $V(r)=r^{2/3}$}
\label{tab:r23}
\begin{center}
\par
\begin{tabular}{cD{.}{.}{50}}
\hline \multicolumn{1}{c}{$(l,\nu)$}&
\multicolumn{1}{c}{$\epsilon_{l\,\nu}$}  \\
\hline
$(0,0)$ &  2.022306599257795366694630473241638339543808208636 \\
$(0,1)$ &  3.06329293436309899696174683955120707141701        \\
$(0,2)$ &  3.83451426291589234862857698779244841              \\
$(0,3)$ &  4.475455263926046368753487903053                   \\
$(0,4)$ &  5.03572773093718648337780666                       \\
$(0,5)$ &  5.539745015667931120170                            \\
$(0,6)$ &  6.00164982551916441                                \\
$(1,0)$ &  2.674632066892448274035790138569168725041769190    \\
$(1,1)$ &  3.5162291388736008172838041603304042376825         \\
$(1,2)$ &  4.1989213934584228295166323216199609               \\
$(1,3)$ &  4.78768391311483216493587370225                    \\
$(1,4)$ &  5.3127607844782485042411545        \\
$(1,5)$ &  5.791063559826215590620            \\
$(1,5)$ &  6.23315893065997295                 \\

\end{tabular}
\par
\end{center}
\end{table}

\begin{table}[tbp]
\caption{Some eigenvalues for the potential $V(r)=r^{3/2}$}
\label{tab:r32}
\begin{center}
\par
\begin{tabular}{cD{.}{.}{30}}
\hline \multicolumn{1}{c}{$(l,\nu)$}&
\multicolumn{1}{c}{$\epsilon_{l\,\nu}$}  \\
\hline
$(0,0)$ &  2.70809241601796914495192943429219    \\
$(0,1)$ &  5.58566253973307550393430185     \\
$(0,2)$ &  8.2268687751240089394177       \\
$(0,3)$ & 10.7317208811602916761         \\
$(0,4)$ & 13.141917795591099             \\
$(1,0)$ &  4.25082600658681113644575609770   \\
$(1,1)$ &  6.9660440200354909840913607     \\
$(1,2)$ &  9.5209043831506977910924       \\
$(1,3)$ & 11.9685512112195160942          \\
$(1,4)$ & 14.336606204183540               \\

\end{tabular}
\par
\end{center}
\end{table}

\end{document}